\documentclass[twocolumn,showpacs,preprintnumbers,amsmath,amssymb,prl]{revtex4}

\usepackage{graphicx}
\usepackage{dcolumn}
\usepackage{bm}
\usepackage{epsfig}
\newcommand{\be}{\begin{eqnarray}}
\newcommand{\ee}{\end{eqnarray}}
\def\refeq#1{(\ref{#1})}

\begin{document}

\title{Magnetic Phase Transitions in One-dimensional Strongly Attractive Three-Component Ultracold Fermions}

\author{ X.W. Guan$^{1}$, M.T. Batchelor$^{1,2}$, C. Lee$^{3}$ and H.-Q. Zhou$^{4}$}

\affiliation{$^{1}$Department of Theoretical Physics, Research
School of Physical Sciences and Engineering, Australian National
University, Canberra ACT 0200, Australia}

\affiliation{$^{2}$Mathematical Sciences Institute, Australian
National University, Canberra ACT 0200, Australia}

\affiliation{$^{3}$Nonlinear Physics Centre and ARC Centre of
Excellence for Quantum-Atom Optics, Research School of Physical
Sciences and Engineering,Australian National University, Canberra
ACT 0200, Australia}

\affiliation{$^{4}$Centre for Modern Physics, Chongqing
University, Chongqing 400044, P.R. China}

\date{\today}

\begin{abstract}

We investigate the nature of trions, pairing and  quantum phase
transitions in one-dimensional strongly attractive  three-component
ultracold fermions in external fields.
Exact results for the groundstate energy, critical fields, magnetization
and phase diagrams are
obtained analytically from the Bethe ansatz solutions.
Driven by  Zeeman splitting, the system shows exotic
phases of trions, bound pairs, a normal Fermi liquid and four mixtures
of these states.
Particularly, a smooth phase transition from a trionic phase into a
pairing phase  occurs as the highest hyperfine level separates from
the two lower energy levels.
In contrast, there is a smooth phase transition from the trionic phase
into a normal Fermi liquid as the lowest level separates from the
two higher levels.

\end{abstract}



\maketitle

There is considerable interest in three-component ultracold
fermions~\cite{Rapp,Lecheminant,Demler,Wilczek}.
Atomic Fermi gases with internal degrees of freedom are
tunable interacting many-body systems featuring novel and subtle
quantum phase transitions.
Two-component Fermi gases of ultracold atoms
with population imbalance have been experimentally observed to
undergo a quantum phase transition between the normal and superfluid
states~\cite{imbalanced-Fermi-gas}.
The bound pairs form a Bardeen-Cooper-Schrieffer (BCS) superfluid,
while the unpaired fermions remain as a separated normal Fermi-Liquid (FL).
These exotic phases have revived interest in the one-dimensional
(1D) integrable model of two-component fermions, which captures the
physics involved in quantum phase transitions and magnetic
ordering~\cite{Orso,Hu,GBLB,Feiguin}.

Three-component fermions reveal more exotic  features
\cite{Rapp,Lecheminant,Demler,Wilczek,Paananen,Rapp2,Thai,Hui}.
The scattering lengths between atoms in different low sublevels
are again tunable via Feshbach resonances \cite{Grimm,Regal,Triples}.
As a consequence, BCS pairing can be favored by anisotropies in
three different ways: specifically, atoms in three low sublevels denoted by
$|1\rangle$, $|2\rangle$ and $|3\rangle$ can form the three possible
pairs $|1\rangle + |2\rangle$, $|2\rangle + |3\rangle$ and
$|1\rangle + |3\rangle$~\cite{Grimm}.
One may also have three internal spin states exhibiting $SU(3)$
symmetry via tuning three scattering lengths close to each other \cite{Luu}.
Significantly, strongly attractive three-component atomic fermions
can form spin-neutral three-body bound states called {\it trions}.
Thus, a phase transition is expected to occur between pairing
superfluid and trionic states~\cite{Rapp,Demler,Rapp2,Wilczek}.

In this Letter, we consider 1D three-component
ultracold fermions with $\delta$-function interaction in  external magnetic fields.
Although this model was solved long ago by the Bethe ansatz (BA)
\cite{Sutherland,Takahashi}, its physics is far from being thoroughly understood.
Here we study the precise nature of trions and pairing in this model,
and calculate critical fields and full phase diagrams by
solving the BA equations and related dressed energy equations.
Our analytical results for magnetism and magnetic-field-driven quantum 
phase transitions in attractive fermions should provide benchmarks for experiments with
ultracold Fermi atoms with multiple internal states.

{\it The model.} The Hamiltonian \cite{Sutherland} we consider
\begin{eqnarray}
{H}&=&-\frac{\hbar ^2}{2m}\sum_{i = 1}^{N}\frac{\partial
^2}{\partial x_i^2}+\,g_{\rm 1D} \sum_{1\leq i<j\leq N} \delta
(x_i-x_j)\nonumber\\
& &+\sum_{i=1}^3N^{i}\epsilon^{i}_Z(\mu_B^{i},B) \label{Ham}
\end{eqnarray}
describes $N$ fermions of mass $m$ and spin-independent s-wave
scattering lengths, which can occupy three possible hyperfine
levels ($|1\rangle$, $|2\rangle$ and $|3\rangle$) and are
constrained to a line of length $L$ with periodic boundary
conditions.
The last term denotes the Zeeman energy, where $N^{i}$ is the
number of fermions in state $| i\rangle$ with Zeeman energy
$\epsilon^{i}_Z$ determined by the magnetic moments $\mu_B^{i}$
and the magnetic field $B$.
The Zeeman energy term can also be expressed as
$ -H_1(N^1-N^2)-H_2(N^2-N^3) +N\bar{\epsilon}$, where the unequally spaced Zeeman
splitting in three hyperfine levels can be characterized by two
independent parameters $H_1 = \bar{\epsilon} -
\epsilon^{1}_Z(\mu_B^{1},B)$ and $H_2 = \epsilon^{3}_Z(\mu_B^{3},B)-
\bar{\epsilon}$, with 
$\bar{\epsilon}=\sum_{\sigma=1}^3\epsilon^{\sigma}_Z(\mu_B^{i},B)/3$
the average Zeeman energy.

In general, the scattering lengths depend on spin states.
However,  it is plausible to
tune three scattering lengths close to each other 
utilizing the broad Feshbach resonances \cite{Grimm,Fermi-1D1}.
Thus the difference in effective interaction parameters
becomes negligible so that three low spin states may have $SU(3)$
degeneracy.
The coupling constant $g_{\rm 1D} =-{\hbar ^2 c}/{m}$ with
interaction strength $c=-{2}/{a_{\rm 1D}}$ determined by the
effective 1D scattering length $a_{\rm 1D}$ \cite{1d-a}.
For simplicity, we define a dimensionless interaction strength
$\gamma=c/n$ with density $n ={N}/{L}$.

The energy eigenspectrum is given in terms of the quasimomenta
$\left\{k_i\right\}$  of the fermions via
$E=\frac{\hbar ^2}{2m}\sum_{j=1}^Nk_j^2$
which in terms of the function $e_n(x)=(x+\mathrm{i}{nc}/{2})/(x-\mathrm{i}{nc}/{2})$
satisfy the nested BA equations \cite{Sutherland,Takahashi}
\begin{eqnarray}
\exp(\mathrm{i}k_jL)&=&\prod^{M_1}_{\ell = 1} e_1\left(
  k_j-\Lambda_\ell \right) \nonumber \\
\prod^N_{\ell = 1}e_1\left(\Lambda_{\alpha}-k_{\ell}
  \right)&=&-\prod^{M_1}_{ \beta =
  1}e_2\left(\Lambda_{\alpha}-\Lambda_{\beta}
  \right)\prod^{M_2}_{\ell=1}e_{-1}\left(\Lambda_{\alpha}- \lambda_{\ell}\right)\nonumber \\
 \prod^{M_1}_{\ell = 1}e_1\left(\lambda_{m}-\Lambda_{\ell}\right)
&=&-\prod^{M_2}_{\ell = 1}e_2\left(\lambda_{m}-\lambda_{\ell}\right)
\label{BE}
\end{eqnarray}
Here $j=1,\ldots, N$, $\alpha = 1,\ldots, M_1$, $m=1,\ldots,M_2$, with
quantum numbers $M_1=N^2+N^3$ and $M_2=N^3$.
The parameters $\left\{\Lambda_{\alpha},\lambda_m\right\}$ are the
rapidities for the internal hyperfine spin degrees of freedom.
For the irreducible representation $\left[3^{N_3}2^{N_2}1^{N_1}
\right]$ three-column Young tableau encode the numbers of
unpaired fermions, bound pairs and trions given by $N_1=N^1-N^2$,
$N_2=N^2-N^3$ and $N_3=N^3$, respectively.

\begin{figure}[t]
{{\includegraphics [width=0.9\linewidth]{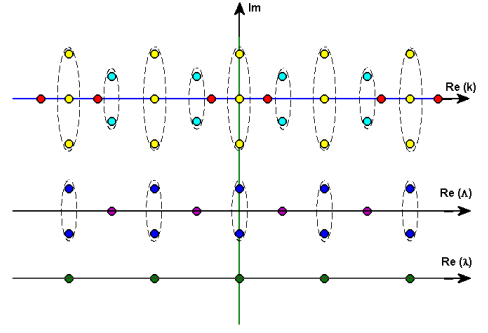}}}
\vspace{-3mm}
\caption{Schematic configuration of Bethe ansatz quasimomenta $k$,
  spin momenta $\Lambda$ and $\lambda$ in the complex plane ($N=29$
  with $N_1=6$, $N_2=4$ and $N_3=5$) at zero temperature.  For
  strongly attractive interaction, the unpaired and paired
  quasimomenta can penetrate into the central region occupied by
  tightly bound trions. }\label{fig:Cartoon}
\end{figure}

{\it Trions and pairing.} In principle, different numbers of
trions, pairs and unpaired fermions can be selected to populate
the groundstate by carefully tuning $H_1$ and $H_2$.
For a state with arbitrary spin polarization, i.e. $N^{i}$ with
$i=1,2,3$ arbitrary, there are
\cite{Takahashi} (i) $N_3$ spin-neutral trions in the quasimomentum
$k$ space accompanied by $N_3$ spin bound states in the
$\Lambda$-parameter space and $N_3$ real roots in the
$\lambda$-parameter space, (ii) $N_2$ BCS bound pairs in $k$ space
accompanied by $N_2$ real roots in $\Lambda$ space and (iii) $N_1$
unpaired fermions in $k$ space.
With the above configuration, we solve the BA equations (\ref{BE}) in the
strong coupling regime $L|c| \gg 1$ to give the quasimomenta for
trions, BCS pairs and unpaired fermions (see Fig.~\ref{fig:Cartoon}), 
from which the energy  is given by
\begin{eqnarray}
\frac{E}{L} &\approx&   \frac{\hbar^2}{2m}
\left[\frac{\pi^2n_1^3}{3}\left(
1+\frac{8n_2+4n_3}{|c|}\right) -\frac{n_2c^2}{2}\right.\nonumber\\
& &\left.+\frac{\pi^2n_2^3}{6}\left(1+\frac{12n_1+6n_2+16n_3}{3|c|}\right)-2n_3c^2\right.\nonumber\\
& &\left.+\frac{\pi^2n_3^3}{9}\left(1+\frac{12n_1+32n_2+18n_3}{9|c|}\right)\right].
\label{E}
\end{eqnarray}
Here $n_a=N_a/L$ with $a=1,2,3$ is the density for unpaired
fermions, pairs and trions, respectively.
This state can be viewed as a mixture of trionic fermions, hard-core bosons
and unpaired fermions, which behave essentially like particles with
different statistical signatures \cite{Lecheminant,GBLB}.
The BCS pair binding energy $\epsilon_{\rm b} = {\hbar^2c^2}/ {(4m)}$ and the
binding energy $\epsilon_{\rm t} = {\hbar^2c^2}/ {m}$ for a trion can be read off Eq. \refeq{E}.
The root patterns obtained from (\ref{BE}) reveal an important
signature, namely unpaired fermions couple to two different kinds --
trionic and BCS pairing -- of FL.
%
%


{\it Thermodynamic Bethe Ansatz.}  In the thermodynamic limit,
i.e. $L,N\to \infty$ with $N/L$  finite, the grand partition function is
$Z=tr(\mathrm{e}^{-\cal{H}/T})=\mathrm{e}^{-G/T}$, where the Gibbs
free energy is $G = E + E_{\rm Z} - \mu N - TS$ in terms of the Zeeman
energy $E_{\rm Z}$, chemical potential $\mu$ and entropy $S$
\cite{Takahashi-B,Schlot,BGOS}.
For finite temperatures, besides complex  BA roots for trions, BCS pairs 
and real roots for unpaired fermions, the quasimomenta $\left\{
\Lambda_{\alpha},\lambda_m \right\}$ form complex strings.
%
%
The Gibbs free energy can be given in terms of the densities of particles
and holes for trions, bound pairs, unpaired fermions as well as spin
degrees of freedom, which are determined from BA (\ref{BE}).
Thus the true physical state is determined by the minimization of the
Gibbs free energy with respect to these densities, which gives rise to a
set of coupled nonlinear integral equations -- the TBA equations
\cite{BG-p}.

Quantum phase transitions in the model 
may be analyzed via the dressed energy equations,
\begin{eqnarray}
\epsilon^{(3)}(\lambda)&=&3\lambda^2-2c^2-3\mu-a_2*{\epsilon^{(1)}}(\lambda) \nonumber\\
& &-\left[a_1+a_3\right]*{\epsilon^{ (2)}}(\lambda)-\left[a_2+a_4\right]*{\epsilon^{ (3)}}(\lambda) \nonumber\\
\epsilon^{(2)}(\Lambda)&=&2\Lambda^2-2\mu-\frac{c^2}{2}-H_2-a_1*{\epsilon^{ (1)}}(\Lambda)  \nonumber\\
& &-a_2*{\epsilon^{2}}(\Lambda) -\left[a_1+a_3\right]*{\epsilon^{(3)}}(\Lambda) \label{TBA-F}\\
\epsilon^{ (1)}(k)&=&k^2-\mu-H_1-a_1*{\epsilon^{ (2)}}(k) -a_2*{\epsilon^{(3)}}(k). \nonumber
\end{eqnarray}
which follow from the TBA equations in the limit $T\to 0$.
Here the function $a_j(x)=\frac{1}{2\pi}\frac{j|c|}{(jc/2)^2+x^2}$
and $\epsilon^{(a)}$ are the dressed energies.
$a_j*{\epsilon^{(a)}}(x)=\int_{-Q_a}^{+Q_a}a_j(x-y){\epsilon^{(a)}}(y)dy$
is the convolution.
The negative part of the dressed energies $\epsilon^{(a)}(x)$ for
$x\le \left|Q_{a}\right|$ correspond to the occupied states in the
Fermi seas of trions, bound pairs and unpaired fermions,
with the positive part of $\epsilon^{(a)}$ corresponding to the unoccupied states.
The integration boundaries $Q_{a}$ characterize the ``Fermi
surfaces'' at $\epsilon^{(a)}(\pm Q_{a})=0$. The zero-temperature
Gibbs free energy per unit length is given by $ G = \sum_{a=1}^3
\frac{a}{2\pi} \int_{-Q_a}^{+Q_a}{\epsilon^{(a)}}(x)dx$. The
chemical potential and magnetization per length are determined by
$H_1$, $H_2$, $g_{1D}$ and $n$ through the relations
\begin{eqnarray}
-\frac{\partial G}{\partial \mu} =n\,,\,-\frac{\partial
G}{\partial H_1}=n_1,\,\, -\frac{\partial
G}{\partial H_2}=n_2, \label{deffield}
\end{eqnarray}

In the absence of analytic solutions of Eq. (\ref{TBA-F}), 
we obtain an exact expansion in the strong coupling regime $\gamma  \gg 1$.
%
Solving the dressed energy equations (\ref{TBA-F}) by iteration
among the relations (\ref{deffield}) and $\epsilon^{\rm a}(\pm
Q_{a})=0$ with $a=1,2,3$, gives the effective chemical potentials
\begin{eqnarray}
\mu^{\rm t} &\approx&
\frac{n_3^2}{9}\left(1+\frac{4n_1}{3|c|}+\frac{32n_2}{9|c|}+\frac{8n_3}{3|c|}\right)+\frac{4n_1^3}{9|c|}+\frac{8n_2^3}{27|c|}
\nonumber\\
\mu^{\rm b} &\approx&\frac{n_2^2}{4}\left(1+\frac{4n_1}{|c|}+\frac{8n_2}{3|c|}+\frac{16n_3}{3|c|}\right)+\frac{4n_1^3}{3|c|}
+\frac{16n_3^3}{81|c|} \nonumber\\
\mu^{\rm u} &\approx& n_1^2\left(1+\frac{8n_2}{|c|}+\frac{4n_3}{|c|}\right)+\frac{2n_2^3}{3|c|}+\frac{4n_3^3}{27|c|}
\label{mu}
\end{eqnarray}
in units of $\hbar^2\pi^2/2m$.
Here we denote $\mu^{\rm t}=\mu+\epsilon_{\rm t}/3$ for trions,
$\mu^{\rm b}=\mu+\epsilon_{\rm b}/2 +H_2/2$ for bound pairs and
$\mu^{\rm u}=\mu+H_1$ for unpaired fermions. These results give
rise to a full characterization of three Fermi surfaces.
It is important to note \cite{BG-p} that the energy for arbitrary
population imbalances can be obtained from $E/L=\mu n+G +n_1
H_1+n_2 H_2$, which coincides with (\ref{E}) derived from the
discrete BA (\ref{BE}). This indicates that the trions and BCS
bound pairs are possibly the true physical states, i.e., the BA
roots comprise the equilibrium states in the thermodynamic limit.

\begin{figure}[h]
{{\includegraphics [width=0.9\linewidth]{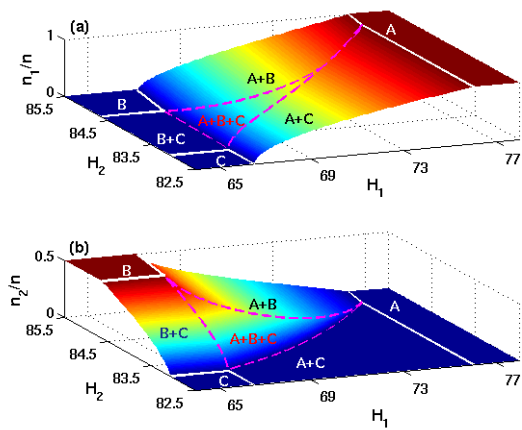}}}
\caption{Phase diagram determined by the energy transfer
  relations (\ref{H1-H2}) with chemical potentials (\ref{mu}) with
  $|c| = 10$ and $n=1$. (a) and (b) show the polarizations
  $n_1/n$ and $n_2/n$ {\em vs} the fields $H_{1}$ and
  $H_{2}$. The figure reveals a novel trion phase $C$, a
  pairing phase $B$, an unpaired phase $A$ and four different mixtures
  of these states.} \label{fig:PD2}
\end{figure}

{\it Full phase diagram.} In the strong coupling regime and in
absence of Zeeman splitting, i.e., $H_1=H_2=0$, the dressed
energies $\epsilon^{(2)}$ and $\epsilon^{(1)}$ are always
positive, i.e., $Q_1=Q_2=0$.
Thus trions form a singlet groundstate.
However, the Zeeman splitting can lift the $SU(3)$
degeneracy and drive the system into different phases.
Breaking a trion state requires a spin excitation energy to diminish an energy gap.
From Eq. (\ref{TBA-F}), the energy transfer relations
among the binding energy, the Zeeman energy and the variation of
chemical potentials between different Fermi seas are given by
\begin{eqnarray}
& &H_1=2c^2/3 +(\mu^{\rm u}-\mu^{\rm t}),\,\, H_2=5c^2/6+2(\mu^{\rm
  b}-\mu^{\rm t}),\nonumber\\
& &H_1-H_2/2= c^2/4+(\mu^{\rm u}-\mu^{\rm b}).
\label{H1-H2}
\end{eqnarray}
These equations determine the full phase diagram and the critical
fields triggered by the Zeeman splitting $H_{1}$ and $H_{2}$.
A similar energy  transfer relation was identified in experiment
for 1D polarized two-component fermions \cite{Fermi-1D1}.
These relations hold for arbitrary interaction strength.
However, for $\gamma \gg 1$, the chemical potentials are given by Eq. (\ref{mu}).
Figure \ref{fig:PD2} shows the polarization, which clearly indicates novel
magnetism, new quantum phases, multicritical points and phase
transitions in terms of Zeeman splitting.

\begin{figure}[t]
{{\includegraphics [width=0.9\linewidth]{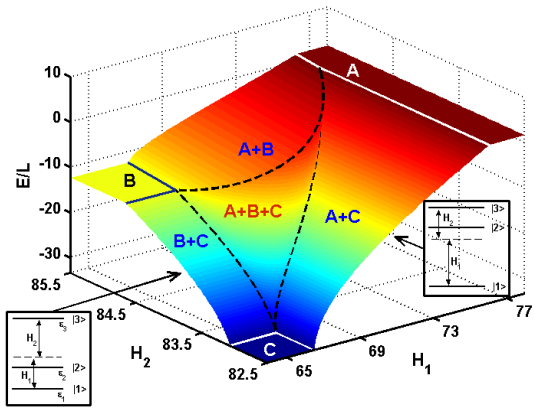}}}
\caption{Groundstate energy {\em vs} Zeeman splitting for $|c| = 10$ and $n=1$. In the vicinity of the
multicritical points and the phase boundaries, the energy surface
varies continuously with $H_1$ and $H_2$. The insets indicate
configurations for the cases of small $H_1$ and small $H_2$.
The dashed lines in the insets  indicate the average Zeeman
energy $\bar{\epsilon}$. }\label{fig:E}
\end{figure}

The groundstate energy {\em vs} Zeeman splitting parameters $H_1$
and $H_2$ can be evaluated from Eq. (\ref{E}) with the
densities $n_1$ and $n_2$ determined from (\ref{H1-H2}). Figure
\ref{fig:E} shows the energy surface for all possible phases shown
in Figure \ref{fig:PD2}.
This figure demonstrates the interplay between different physical groundstates.
We see  a mixture  of trions, BCS pairs and unpaired fermions
($A+B+C$) populates the groundstate for  certain values of $H_1$ and $H_2$.
There are six different phase transitions across the
$A+B+C$ boundaries in the $H_1$--$H_2$ plane.
All phase transitions between two phases are second order and reveal
a universality class of linear-field-dependent magnetization with
finitely divergent susceptibility in the vicinities of critical fields.
Our analytical results (\ref{H1-H2}) do not support the
square-root-field-dependent behaviour of magnetization argued in
\cite{Schlot}, but do agree with results for the attractive Hubbard
model \cite{Penc}.
%
%

For small $H_1$ (i.e., small splitting between the two lower levels), a
smooth phase transition from a trionic state into a mixture of trions
and pairs occurs as $H_2$ exceeds the lower critical value $H_2^{c1}$ (see Fig. \ref{fig:E}).
%
%
When $H_2$ is greater than the upper critical value $H_2^{c2}$, the
atoms within the two lower states form a pure  pairing phase with $SU(2)$
symmetry and the highest level remains unpopulated.
In this pure pairing phase, the three-level system is reduced to a
two-level one.
Trions and BCS pairs coexist when $H_{2}^{c1}<H_2<H_{2}^{c2}$.
The critical fields
$H_{2}^{c1} \approx \frac{\hbar^2n^2}{2m} \left(
\frac{5\gamma^2}{6} - \frac{2\pi^2}{81}(1+\frac{8}{27|\gamma|})
\right)$ and $H_{2}^{c2} \approx \frac{\hbar^2n^2}{2m} \left(
\frac{5\gamma^2}{6} + \frac{\pi^2}{8}(1+\frac{20}{27|\gamma|}) \right)$
are uniquely determined by the second equation in (\ref{H1-H2}).
The polarization curve $n_2/n$ indicates that the phase
transitions in the vicinities of the critical lines $H_{2}^{c1}$
and $H_{2}^{c2}$ are of second order.
In addition, the phase transitions $B \rightarrow A+B \rightarrow
A $ induced by increasing $H_1$ are reminiscent of those in the
two-component systems \cite{GBLB,Penc}.
We see clearly that equally spaced Zeeman splitting $H_1=H_2$ does not favour 
spin-dependent charge states in 1D multi-component attractive fermions.

For small $H_2$ (i.e., small splitting between the two higher levels),
the pairing phase is not favored and the trions are 
broken into unpaired fermions in the lowest level (see Fig.~\ref{fig:E}).
Using the first relation in (\ref{H1-H2}), we see that
the trionic state with zero polarization $n_1/n = 0$ forms
the groundstate when the field $H<H_{1}^{c1}$. Here $H_{1}^{c1}
\approx \frac{\hbar^2n^2}{2m} \left( \frac{2\gamma^2}{3} -
\frac{\pi^2}{81}(1+\frac{4}{9|\gamma|}) \right)$ is the lowest
critical field which makes the excitation gapless.
If the lowest level is widely separated from the two higher
levels, i.e., $H_1>H_{1}^{c2} \approx \frac{\hbar^2n^2}{2m} \left(
\frac{2\gamma^2}{3} + \pi^2(1-\frac{4}{9|\gamma|}) \right)$, all
trions are broken and the state becomes a normal FL where all
atoms occupy the lowest level.
Trions and unpaired fermions coexist in the 
intermediate region $H_{1}^{c1}<H_1<H_{1}^{c2}$.

For trapped systems, these exotic phases and their
segments in real space can be predicted with the help of the local density
approximation.
For example, for small $H_1$, a mixture of trions and BCS pairs lies
in the trap centre and the fully trionic phase (or the fully
pairing phase) sits in the two outer wings for $H_2 <
\frac{5\hbar^2n^2\gamma^2}{12m}$ (or $H_2 >
\frac{5\hbar^2n^2\gamma^2}{12m} $).
In contrast, for small $H_2$, a mixture of trions and unpaired
fermions lies in the centre and the fully trionic phase (or the
fully unpaired phase) sits in the two outer wings for $H <
\epsilon_{\rm t}/3$ (or $H > \epsilon_{\rm t}/3$).

To conclude, our analytic BA and TBA results for the critical fields,
quantum phase transitions and full phase diagrams reveal the nature of
spin-neutral trions and BCS pairing in three-component ultracold
fermions in external fields.
Particularly, we found transitions between trionic and
BCS pairing phases and between trionic and normal FL phases may
occur for certain types of Zeeman splitting.
These exotic phases should stimulate further experimental interest
in multi-component ultracold Fermi gases with mismatched Fermi surfaces.

This work has been supported by the Australian Research Council.
The authors thank  N. Andrei, M.A. Cazalilla, H. Frahm, W.V. Liu
and M. Takahashi for helpful discussions. C.L. thanks Yu.S. 
Kivshar for support.

\end{document}